\documentclass[pdftex,twocolumn,epjc3]{svjour3}          

\RequirePackage[T1]{fontenc}

\smartqed  

\RequirePackage{graphicx}
\RequirePackage{mathptmx}      
\RequirePackage{flushend}
\RequirePackage[numbers,sort&compress]{natbib}
\RequirePackage[colorlinks,citecolor=blue,urlcolor=blue,linkcolor=blue]{hyperref}
\usepackage{subfigure}
\usepackage{color}
\usepackage{amssymb}
\usepackage{amsmath}
\usepackage{balance}
\usepackage{mathtools}

\newcommand{\sech}{\normalfont\mbox{sech}\,}

\journalname{Eur. Phys. J. B}

\begin{document}

\title{Nonlinear tunneling of solitons in a variable coefficients nonlinear Schr\"odinger equation with $\mathcal{PT}$-symmetric Rosen-Morse potential}

\author{K.~Manikandan \thanksref{addr1}, J.B. Sudharsan \thanksref{addr2} \and M. Senthilvelan \thanksref{e1,addr1} }
\thankstext{e1}{e-mail: velan@cnld.bdu.ac.in}

	\institute{Department of Nonlinear Dynamics, Bharathidasan University, Tiruchirappalli 620024, Tamil Nadu, India \label{addr1}. \and SSN Research Centre, SSN College of Engineering, Kalavakkam, Chennai - 603 110, Tamil Nadu, India \label{addr2}.}

\abstractdc{
We construct soliton solution of a variable coefficients nonlinear Schr\"odinger equation in the presence of parity reflection - time reversal $(\mathcal{PT})-$ symmetric Rosen-Morse potential using similarity transformation technique.  We transform the variable coefficients nonlinear Schr\"odinger equation into the nonlinear Schr\"odinger equation with $\mathcal{PT}-$symmetric potential with certain integrability conditions. We investigate in-detail the features of the obtained soliton solutions with two different forms of dispersion parameters.  Further, we analyze the nonlinear tunneling effect of soliton profiles by considering two different forms of nonlinear barrier/well and dispersion barrier/well.  Our results show that the soliton can tunnel through nonlinear barrier/well and dispersion barrier/well with enlarged and suppressed amplitudes depending on the sign of the height. Our theoretical findings are experimentally realizable and might help to model the optical devices.}
	
\maketitle

\section{Introduction}

In recent years, many researchers from physics and engineering community turned their attention into optical solitons by considering wide applications ranging from long distance communication to ultra fast optical switching devices \cite{1a,1b,1,2,3,4}. Hasegawa and Tappert \cite{5} theoretically proposed the concept of optical solitons in a  fiber and it was later experimentally realized by Mollenauer et al. \cite{6}.  In real time communication applications, the optical pulses can propagate to very long distances if one can have control over the width of the pulses.

In theory, the nonlinear Schr\"odinger equation (NLS) with constant coefficients have limited control over the propagation of optical pulses. On the other hand, inhomogeneous NLS equation with different nonlinearities may offer effective control over the propagating optical pulses with suitable dispersion, nonlinearity, gain and loss parameters all over the propagation distance \cite{7,8}. When a system contains gain and loss distribution, soliton can exist only for  certain specific values of propagation constant. Since the complex $\mathcal{PT}-$ symmetric potential in optical system can admit real linear spectra the soliton could exist at continuous ranges \cite{9,10,11}.

For the past two decades, the concept of $\mathcal{PT}$ symmetry has drawn plethora of interest from both theoretical and experimental perspective \cite{12}. Even though, a complex $\mathcal{PT}$ symmetric potential makes a simple Hamiltonian as non-Hermitian in nature, it may support a real eigenvalued spectra \cite{12,13,14}. For such a complex $\mathcal{PT}$ symmetric potential, the real part should be symmetric function of position whereas imaginary part should be anti-symmetric function of position. Particularly in wave guided optics, the refractive index ($n(x)$) of the crystal fiber can be regarded as complex $\mathcal{PT}$ symmetric potential which satisfies the condition $n(x) = n^*(-x)$. Actually, the real and imaginary part of $n(x)$ describes the refractive index profile and the gain (+) or loss (-) respectively. 

The presence of complex $\mathcal{PT}$ symmetric potential in a variable coefficient NLS equation plays a key role in nonlinear soliton tunneling effect. At first, Serkin and Belyaeva \cite{15} studied the tunneling of solitons in a variable coefficients NLS equation. Their work initiated the new and exciting application of soliton tunneling in optical pulse propagation. For example, the compression of optical soliton in the presence of nonlinear barrier was studied by Yang et.al. \cite{16}. Zhong and Belic \cite{17} have studied the effects of nonlinear tunneling of spatial solitons in diffractive and nonlinear barriers.  The tunneling effects of bright and dark similaritons in a generalized coupled NLS equation were studied in Refs. \cite{18,18a}. To the best of our knowledge, no studies have been carried out in optical soliton tunneling with a complex Rosen-Morse $\mathcal{PT}$ symmetric potential. Hence, in this paper, we study the nonlinear tunneling effects of optical pulse with $\mathcal{PT}$ symmetric Rosen-Morse potential by varying different parameters that appear in the variable coefficients NLS equation.

In \cite{15,19,20,21,22,22a,22b}, the dynamics of non-autonomous solitons like width variation and energy control are discussed by varying the dispersion, nonlinearity and gain/loss parameters. The propagation of non-autonomous solitons and tunneling in a optical fiber are discussed in \cite{17,22}. In \cite{23}, the authors have investigated the impact of complex $\mathcal{PT}$ symmetric potential in defocusing Kerr media on the dynamics of spatial bright solitons.  Stability of dark solitons and the vortex dynamics in the complex $\mathcal{PT}$ symmetric nonlinear media was reported in \cite{24}. Stabilization and dynamics of higher dimensional solitons in inhomogeneous NLS equation with complex $\mathcal{PT}$ symmetric potential was studied in \cite{25}. A new kind of dynamics was observed in optical solitons with complex $\mathcal{PT}$ symmetry in \cite{26,27,28,28a} and with periodic complex $\mathcal{PT}$ symmetric potential in \cite{29}.

In this work, we intend to explore the dynamical behaviours of inhomogeneous solitons with the presence of Rosen-Morse $\mathcal{PT}$ - symmetric potential.  For this study, we first construct soliton solutions for the NLS equation with distributed coefficients, namely dispersion, nonlinearity, tapering and $\mathcal{PT}$-symmetric potential parameters through the similarity transformation technique.  We consider two different forms of dispersion parameter, namely constant and exponentially distributed dispersion parameter. Our results show that for the constant dispersion parameter, the soliton structure reveals the habitual feature whereas when the dispersion parameter is exponentially distributed, the amplitude of soliton monotonically enhances along the propagation distance in the $x-z$ plane.  Further, we analyze how the optical soliton profiles get modified by varying the parameters including the strengths of real and imaginary part of potentials, nonlinearity and dispersion decaying or enhancing parameter.  In addition to the above, we investigate the tunneling effects of optical soliton by propagating it through two different types of nonlinear barrier/well and dispersion barrier/well with and without the exponential background.

We organize our work as follows.  In Sec. 2, we construct the soliton solutions of variable coefficients NLS equation with Rosen-Morse $\mathcal{PT}$-symmetric potentials. By imposing proper integrability conditions, the system under investigation is transformed into constant coefficient NLS equation with complex $\mathcal{PT}$-symmetric potential.  In Sec. 3, by taking two different forms of dispersion parameters, we investiage the dynamical features of the constructed soliton profiles. In Sec. 4, we study the nonlinear tunneling effects of soliton profiles by considering two different forms of nonlinear barrier/well and dispersion barrier/well. Finally, in Sec. 4, we summarize our findings. 
\section{Model and Similarity transformation}
We consider the following (1+1)-dimensional NLS equation with distributed coefficients \cite{1,22,22a} 
\begin{align}
i\psi_{z}+\frac{\beta(z)}{2}\psi_{xx}+R(z) \vert \psi \vert^2 \psi +\alpha(z) x^2 \psi+ [v(x,z)\nonumber \\ +i w(x,z)]\psi=0, 
\label{cdis:eq1}
\end{align}
where $\psi(x,z)$ represents complex envelope of the optical field, $z$ denotes the dimensionless coordinate along the direction of propagation and $x$ is the spatial variable.  The parameters $\beta(z)$ and $R(z)$ denote the coefficients of dispersion and nonlinearity, respectively. With respect to the focusing $(R>0)$ or defocusing $(R<0)$ nature of the optical medium, the tapering function (dimensionless) $\alpha(z)$ can either be negative or positive.  The even real part function $(v(x,z))$ and odd imaginary part function $(w(x,z))$ of optical potential represents the index guiding and gain or loss distribution, respectively.  The functions, $\beta(z)$, $R(z)$ and $\alpha(z)$ are real analytic functions of $z$. 

To derive inhomogeneous soliton solutions of (\ref{cdis:eq1}), we consider the transformation in the form
\begin{equation}
\psi(x,z)=\rho(z)U(\xi(x,z),\tau(z))\exp[i \phi(x,z)],
\label{cdis:eq2}
\end{equation}
where $U(\xi(x,z),\tau(z))$ is the solution of the partial differential equation of the form
\begin{equation}
i U_{\tau}+\frac{U_{\xi\xi}}{2}+|U|^2U+[V(\xi)+i W(\xi)]U=0.
\label{nlspt}
\end{equation}
In this transformation (\ref{cdis:eq2}),  the amplitude $\rho(z)$,  the effective propagation distance $\tau(z)$,  the similarity variable $\xi(x,z)$ and the phase $\phi(x,z)$ are all to be determined.  Substituting the transformation (\ref{nlspt}) into (\ref{cdis:eq1}), we find 
\begin{align}
\label{pdes}
&& \xi_{xx}  = 0, \notag \\
&& \tau_z-R(z)\rho^2(z) = 0, \notag \\
&& \beta(z)\xi_x^2-2 R(z)\rho^2(z) = 0, \notag \\
&& 2\rho(z)+\beta(z)\phi_{xx}\rho(z) = 0, \notag \\
&& \xi_z+\beta(z)\xi_x\phi_x = 0, \notag \\
&& 2 \alpha(z) x^2-2\phi_z-\beta(z)\phi_x^2 = 0, \notag \\
&& v(x,z) = \rho^2(z)R(z)V(\xi), \notag \\
&& w(x,z) = \rho^2(z)R(z)W(\xi).  
\end{align}

Solving these partial differential equations, we obtain the following expressions, namely  
\begin{subequations}
	\begin{align}
	\label{cdis:eq3a} \xi(x,z) = & \, \frac{x}{m(z)}, \;\; m(z)=\frac{\beta(z)}{2R(z)}, \;\; \rho(z)= \frac{1}{\sqrt{m(z)}}, \\
	\label{cdis:eq3b} \phi(x,z)=& \, \frac{m_zx^2}{2\beta(z)m(z)}, \;\; \tau(z)= 2 \int \frac{R^2(z)}{\beta(z)}dz.  
	\end{align}
\end{subequations}

Equation (\ref{nlspt}) admits solitary wave solutions for different forms of $\mathcal{PT}$ - symmetric potentials, see for example Refs. \cite{22,25}.  A number of works have been devoted to study the characteristics of inhomogeneous solitons by considering different kinds of $\mathcal{PT}$-symmetric potentials including Scarff-II and optical lattice potential.  In our study, we consider the Rosen-Morse $\mathcal{PT}-$symmetric potential \cite{30,31}.  With $V(\xi)= -a(a+1) \sech^2(\xi)$ and $W(\xi)= 2 b \tanh(\xi)$, Eq. (\ref{nlspt}) yields the solitary wave solution of the form \cite{30}
\begin{align}
\label{soln}
U(\xi,\tau)=\sqrt{a^2+a+1} \; \sech{(\xi)}\exp[i(b\xi+\lambda \tau)],
\end{align} 
where $a$ and $b$ are the real and imaginary strengths of Rosen-Morse $\mathcal{PT}-$symmetric potential.  

One can identify the localized solutions in (\ref{cdis:eq1}) provided the following constraints are fulfilled, namely
\begin{align}
\label{constr1}
\displaystyle \alpha(z)=\dfrac{\splitdfrac{\beta^2(z)\left(2R^2_z-R(z)R_{zz}\right)-R^2(z)\beta^2(z)}{ +R(z)\beta(z)\left(-R_z\beta_z+R_z\beta_{zz}\right)}}{2\beta^2(z)\beta^3(z)},
\end{align}
and 
\begin{align}
\label{constr2}
v(x,z)=\frac{\beta(z)V(\xi)}{R^2(z)}, \;\;\; w(x,z)=\frac{\beta(z)W(\xi)}{R^2(z)}.
\end{align}
We note here that the conditions (\ref{constr2}) are extra constraints which arise due to the presence of PT-symmetric potential.  These constraints do not appear when one transforms the inhomogeneous NLS equation to homogeneous NLS equation.  
\begin{figure*}[!ht]
	\begin{center}
		\subfigure[]
		{
			\resizebox{0.4\textwidth}{!}{\includegraphics{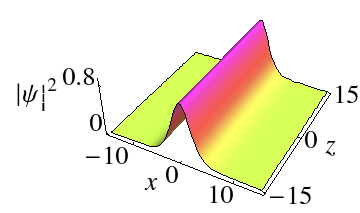}}
			\label{fig:1}
		}~~ 
		\subfigure[]
		{
			\resizebox{0.4\textwidth}{!}{\includegraphics{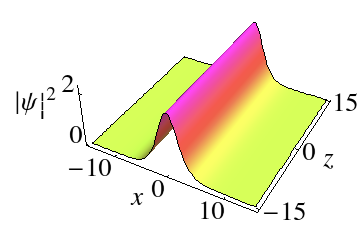}}
			\label{fig:2}
		}\\ \vspace{-0.3cm}
		\subfigure[]
		{
			\resizebox{0.4\textwidth}{!}{\includegraphics{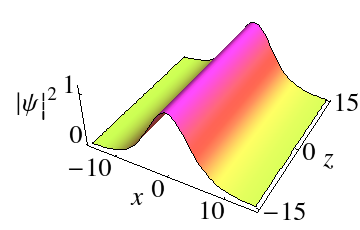}}
			\label{fig:3}
		}~~
		\subfigure[]
		{
			\resizebox{0.4\textwidth}{!}{\includegraphics{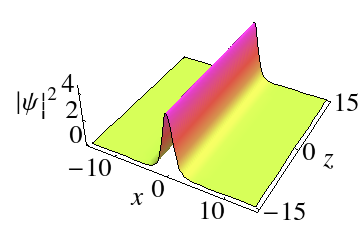}}
			\label{fig:4}
		}
	\end{center}
	\vspace{-0.3cm}
	\caption{Intensity profile of soliton of (\ref{cdis:eq1}) for constant dispersion parameter $\beta(z)=\beta_0$ with \textbf{(a)} $a=0.25$, $b=0.5$, $\beta_{0}=0.05$ and $R_0=0.01$; \textbf{(b)} $a=1.5$, $b=1.25$, $\beta_0=0.05$ and $R_0=0.01$; \textbf{(c)} $a=1.5$, $b=1.25$, $\beta_{0}=0.1$ and $R_0=0.01$; \textbf{(d)} $a=1.5$, $b=1.25$, $\beta_{0}=0.05$ and $R_0=0.02$.}
	\label{fig1}
\end{figure*}
\begin{figure*}[!ht]
	\begin{center}
		\subfigure[]
		{
			\resizebox{0.4\textwidth}{!}{\includegraphics{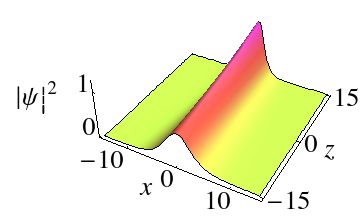}}
			\label{fig:1}
		}~~ 
		\subfigure[]
		{
			\resizebox{0.4\textwidth}{!}{\includegraphics{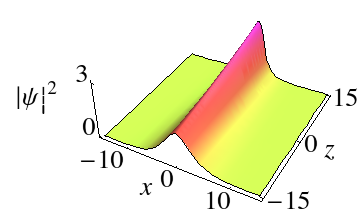}}
			\label{fig:2}
		}\\ \vspace{-0.3cm}
		\subfigure[]
		{
			\resizebox{0.4\textwidth}{!}{\includegraphics{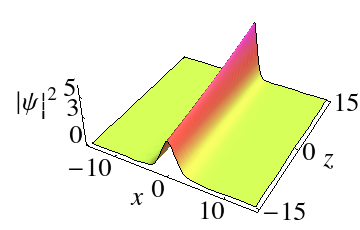}}
			\label{fig:3}
		}~~
		\subfigure[]
		{
			\resizebox{0.4\textwidth}{!}{\includegraphics{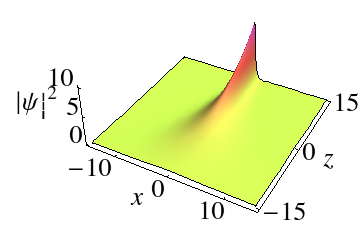}}
			\label{fig:4}
		}
	\end{center}
	\vspace{-0.3cm}
	\caption{Intensity profile of soliton of (\ref{cdis:eq1}) for an exponentially distributed dispersion parameter $\beta(z)=\beta_0$ with \textbf{(a)} $a=0.25$, $b=0.5$, $\beta_{0}=0.05$, $R_0=0.01$ and $\sigma=0.02$; \textbf{(b)} $a=1.5$, $b=1.25$, $\beta_0=0.05$, $R_0=0.01$ and $\sigma=0.02$; \textbf{(c)} $a=1.5$,$b=1.25$, $\beta_{0}=0.1$, $R_0=0.02$ and $\sigma=0.02$; \textbf{(d)} $a=1.5$, $b=1.25$, $\beta_{0}=0.1$, $R_0=0.02$ and $\sigma=0.1$.}
	\label{fig2}
\end{figure*}
The exact solution of Eq. (\ref{cdis:eq1}) can be obtained by properly choosing the suitable forms of $\beta(z)$ and $R(z)$ with the above constraints in the following form
\begin{align}
\label{cdis:eq6}
\psi(x,z)&=& \sqrt{\frac{2R(z)}{\beta(z)}}U(\xi,\tau)\exp\left[i\left(\frac{(R(z)\beta_z-\beta(z)R_z)x^2}{2\beta^2(z)R(z)}\right)\right],
\end{align}
where $U(\xi,\tau)$ is given in Eq. (\ref{soln}). By suitably selecting arbitrary functions, we can identify certain optical soliton solutions of (\ref{cdis:eq1}). 
\section{Dynamical characteristics of inhomogeneous solitons}
To analyse and understand the behaviour of exact soliton solutions (\ref{cdis:eq6}) of (\ref{cdis:eq1}), two different forms of dispersion parameters are considered. In the following, we study the characteristics of inhomogeneous solitons for two different forms of dispersion parameters.

\subsection{Case 1}
We start our investigation by fixing the dispersion parameter as a constant, $\beta(z)=\beta_0$.  Substituting this dispersion parameter and the noninearity parameter $R(z)=R_0$ into (\ref{cdis:eq6}), we can obtain the soliton solutions of (\ref{cdis:eq1}).  

\begin{figure*}[!ht]
	\begin{center}
		\subfigure[]
		{
			\resizebox{0.4\textwidth}{!}{\includegraphics{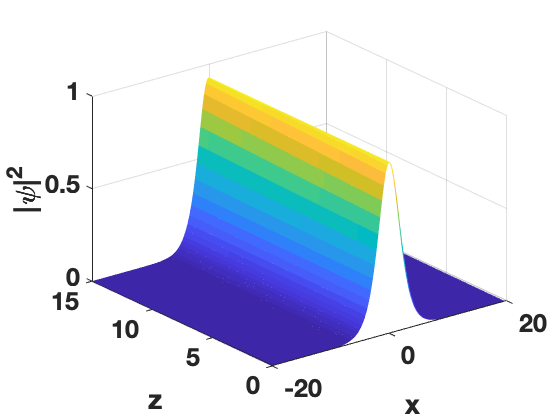}}
			\label{fig:1}
		}~~ 
		\subfigure[]
		{
			\resizebox{0.4\textwidth}{!}{\includegraphics{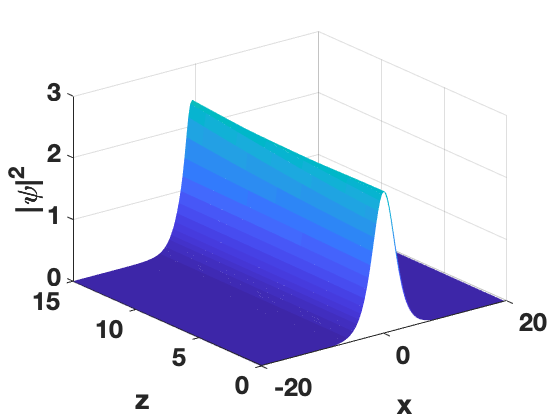}}
			\label{fig:2}
		}\\ \vspace{-0.3cm}
		\subfigure[]
		{
			\resizebox{0.4\textwidth}{!}{\includegraphics{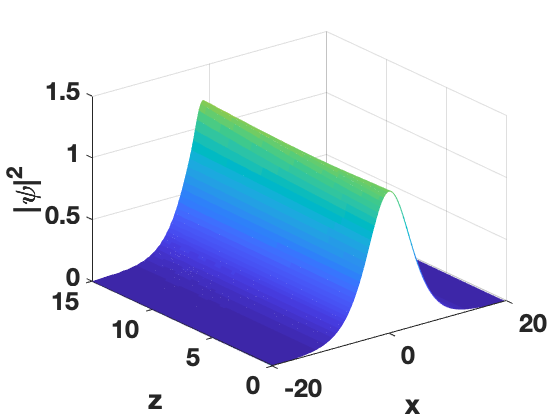}}
			\label{fig:3}
		}~~
		\subfigure[]
		{
			\resizebox{0.4\textwidth}{!}{\includegraphics{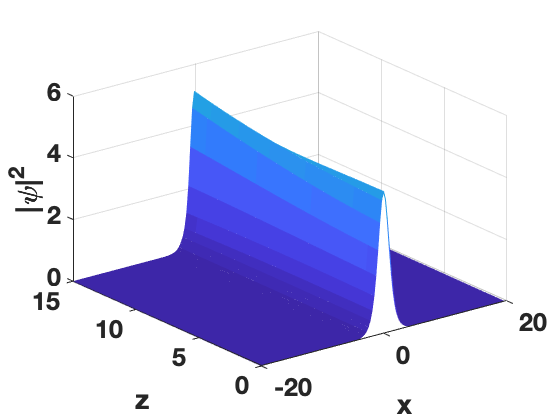}}
			\label{fig:4}
		}
	\end{center}
	\vspace{-0.3cm}
	\caption{Numerically obtained soliton profile of (\ref{cdis:eq1}) for $\beta(z)=\beta_0$ with the initial solution (\ref{cdis:eq6}). The other parameters are same as in Fig. \ref{fig1}. }
	\label{fig3}
\end{figure*}
\begin{figure*}[!ht]
	\begin{center}
		\subfigure[]
		{
			\resizebox{0.4\textwidth}{!}{\includegraphics{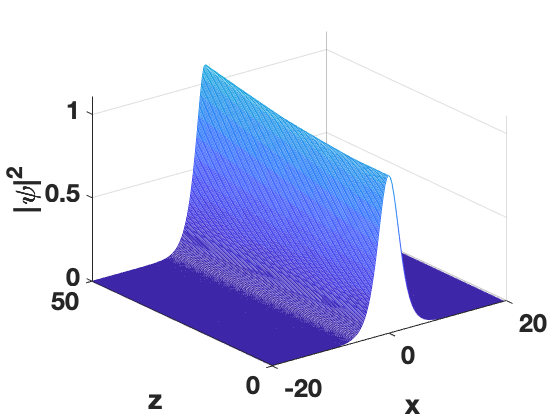}}
			\label{fig:1}
		}~~ 
		\subfigure[]
		{
			\resizebox{0.4\textwidth}{!}{\includegraphics{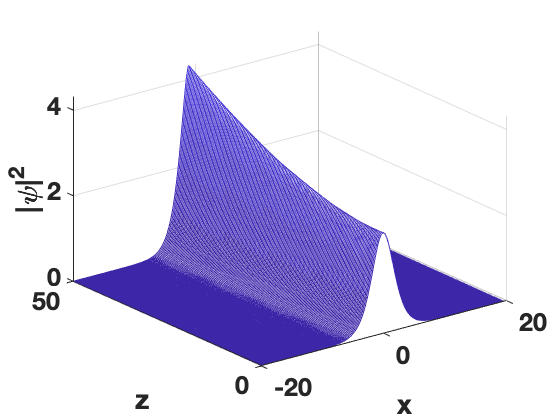}}
			\label{fig:2}
		}\\ \vspace{-0.3cm}
		\subfigure[]
		{
			\resizebox{0.4\textwidth}{!}{\includegraphics{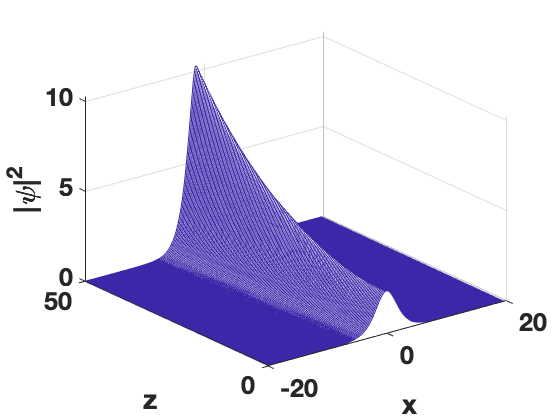}}
			\label{fig:3}
		}~~
		\subfigure[]
		{
			\resizebox{0.4\textwidth}{!}{\includegraphics{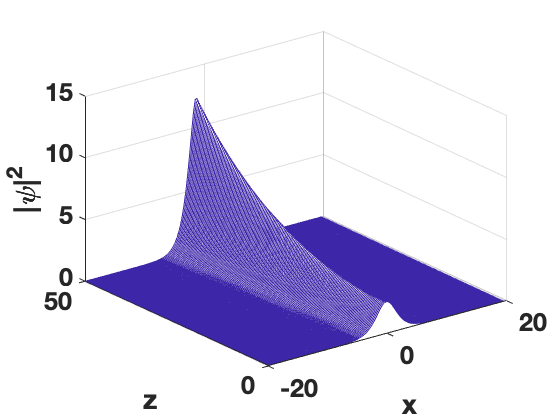}}
			\label{fig:4}
		}
	\end{center}
	\vspace{-0.3cm}
	\caption{Numerically obtained soliton profile of (\ref{cdis:eq1}) for $\beta(z)=\beta_0 \exp{(-\sigma z)}$ with initial solution (\ref{cdis:eq6}).  The other parameters are same as in Fig. \ref{fig2}.}
	\label{fig4}
\end{figure*}

Figure \ref{fig1} displays the intensity profiles of solitons with the presence of $\mathcal{PT}$ - symmetric Rosen-Morse potential for the constant dispersion parameter as a function of different parameters, such as the strength of dispersion parameter ($\beta_0$), nonlinearity parameter ($R_0$), real and imaginary parts of Rosen-Morse $\mathcal{PT}-$symmetric potentials ($a$ and $b$), respectively.  In Fig. \ref{fig1}(a) we show the soliton profile for $a=0.25$, $b=0.5$, $\beta_{0}=0.05$ and $R_0=0.01$.  We observe that the soliton profile reveals the fundamental feature, that is soliton propagates with constant velocity.  When we increase the strength of real and imaginary parts of $\mathcal{PT}-$symmetric Rosen-Morse potential in the obtained soliton solutions with $\beta_{0}=0.05$, $R_0=0.01$, we see the enhanced intensity soliton profile which is represented in Fig. \ref{fig1}(b).  We then vary the strength of dispersion parameter $\beta_0$ to $0.1$ and fix the other parameters similar to the previous case.  The resultant profile of this case is shown in Fig. \ref{fig1}(c).  The outcome reveals that the amplitude of the soliton is decreasing and the width of the soliton is increasing.  Next, we move on to investigate the role of nonlinearity parameter strength $R_0$.  By tuning $R_0$ to $0.02$, the enhancement of amplitude and compression of width of the soliton are noticed as clearly demonstrated in Fig. \ref{fig1}(d).  
\subsection{Case 2}
Next we study the impact of an exponentially distributed dispersion parameter, $\beta(z)=\beta_0 \exp(-\sigma z)$ with $R(z)=R_0$ on the soliton profiles.  Here $\sigma > 0$ and $\sigma<0$ describe the dispersion decreasing fiber and the dispersion enhancing fiber, respectively. Exponentially distributed system offers myriad applications in long-distance optical communication systems \cite{7}.

The qualitative nature of intensity profiles of solitons with the presence of $\mathcal{PT}-$symmetric Rosen-Morse  potential for an exponentially distributed dispersion parameter with variation to other parameters are presented in Figs. \ref{fig2}(a)-(d).   Here also we analyze the impact of different strengths of the physical parameters on the obtained soliton profiles.  For the choice $a=0.25$, $b=0.5$, $\beta_{0}=0.05$, $R_0=0.01$ and $\sigma=0.02$,  we visualize a soliton profile as shown in \ref{fig2}(a).  Here we notice that the amplitude of soliton increases along the direction of propagation in the $x-z$ plane.  When we vary the strength of real and imaginary parts of $\mathcal{PT}-$symmetric Rosen-Morse potential to $a=1.5$ and $b=1.25$ while keeping the other parameters same as in  the above choice, we observe that the intensity of soliton raises as displayed in Fig. \ref{fig2}(b).  In Fig. \ref{fig2}(c), we show the enhanced soliton profile which is obtained by changing the parameter $R_0$ to $0.02$.  When the dispersion decaying/increasing parameter ($\sigma$) is varied from $0.02$ to $0.1$, we observe that the soliton disappears in the regime $z<0$, re-emerges at $z=0$ and propagates thereafter with increasing amplitude which is demonstrated in Fig.  \ref{fig2}(d).  From these observations, we notice that in all the four considerations the amplitude of soliton monotonically increases along the direction of propagation. Thus, in the absence of other relay devices one can amplify the intensity of solitons by exploiting the exponentially distributed dispersion parameter or in other words the inhomogeneous nature of optical fibers.

Next we perform a direct numerical simulation of (\ref{cdis:eq1}) with (\ref{cdis:eq6}) as the initial input by employing well known split-step Fourier method \cite{32}.  We consider the space $(x)$ range from -20 to +20 and the propagation distance $(z)$ range from 0 to 50. We take the space step $\delta x = 0.3$ and the time step $\delta t = 0.006$. We have also checked the compatibility of the obtained numerical results by changing both space and time steps.  We numerically generate the intensity of soliton profiles for the constant dispersion coefficient with the parameter values mentioned in Fig. \ref{fig1} and present the outcome in Figs. \ref{fig3}(a)-(d). Similarly in Fig. \ref{fig4}(a)-(d), we numerically show the amplification of soliton profiles for the case of exponentially distributed dispersion coefficient with the parameter choices as mentioned in Fig. \ref{fig2}. There is a very good agreement between the numerical results and the analytical predictions. Thus, our numerical simulation further confirms the existence of the soliton in the considered system.

\section{Nonlinear tunneling effect}
In this section, we study the nonlinear tunneling effect of soliton solutions of (\ref{cdis:eq1}).  In our investigations, we consider two different forms of nonlinear barrier/well and two types of dispersion barrier/well.  
\subsection{Nonlinear barrier/well}
To explore the performance of optical soliton through nonlinear tunneling effect, we consider two cases, namely nonlinear barrier/well with and without an exponential background.  More specifically, we intend to analyze how the soliton profiles get affected at the barrier/well.
\subsubsection{Nonlinear tunneling without exponential background}
To analyze the tunneling effects of the obtained optical soliton solutions, in the following, we consider nonlinear barrier/well without exponential background of the form
\begin{equation}
\label{barrier1}
\beta(z)=\beta_0, \;\;\; R(z)= R_0+h \; \sech^2{[\kappa (z-z_0)]},
\end{equation}
where $h$, $\kappa$ and $z_0$ represents the height, width and location of the nonlinear barrier/well, respectively.  With the positive sign of $h$ the nonlinear form $R(z)$ acts as nonlinear barrier and with the negative sign of $h$, $R(z)$ acts as nonlinear well.  
\begin{figure*}[!ht]
	\begin{center}
		\subfigure[]
		{
			\resizebox{0.4\textwidth}{!}{\includegraphics{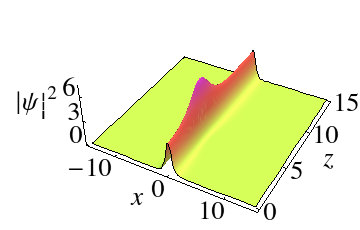}}
			\label{fig:1}
		}~~ 
		\subfigure[]
		{
			\resizebox{0.4\textwidth}{!}{\includegraphics{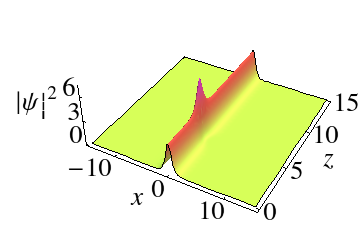}}
			\label{fig:2}
		}\\ \vspace{-0.3cm}
		\subfigure[]
		{
			\resizebox{0.4\textwidth}{!}{\includegraphics{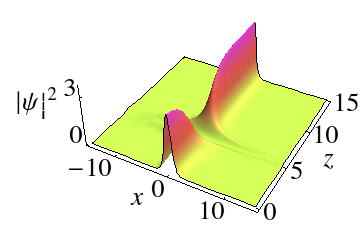}}
			\label{fig:3}
		}~~
		\subfigure[]
		{
			\resizebox{0.4\textwidth}{!}{\includegraphics{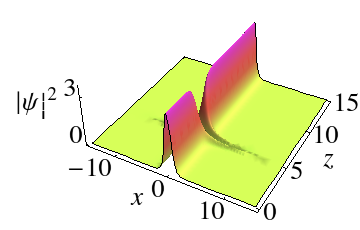}}
			\label{fig:4}
		}
	\end{center}
	\vspace{-0.3cm}
	\caption{Propagation of soliton through nonlinear barrier (a) and (b) for $h=1$ and nonlinear well (c) and (d) for $h=-1$.  The other parameters are $a=0.5,b=0.75, \beta_0=1.5$, $\chi_0=1$, $z_0=5$ with (a)  and (c) $\kappa=0.5$ and (b) and (d) $\kappa=1.5$. }
	\label{fig5}
\end{figure*}

Figure \ref{fig5} represents the dynamics of the nonlinear tunneling effect of the obtained optical soliton (\ref{cdis:eq6}) propagating through the nonlinear barrier/well (\ref{barrier1}).  For $h=1$ in (\ref{barrier1}), we observe that the optical soliton tunnels through nonlinear barrier which is demonstrated in Figs. \ref{fig5}(a)-(b).  The barrier is formed at $z_0=5$.  In Fig. \ref{fig5}(a), we observe that when the soliton propagates through the barrier, its amplitude enhances at $z=z_0=5$ along the direction of propagation.  After reaching the barrier, the soliton propagates with its initial profile.  We notice that width of the soliton elongates along the direction of propagation at the barrier for a lower value of barrier's width ($\kappa=0.5$).  When we increase the value of $\kappa$ to $1.5$, width of the soliton gets compressed at the barrier.  This outcome is demonstrated in Fig. \ref{fig5}(b).  For the choice $h=-1$, we can visualize the propagation of soliton through nonlinear well which is presented in Figs. \ref{fig5}(c)-(d).  From fig. \ref{fig5}(c), we notice that when the soliton propagates through the well, amplitude of the pulse diminishes and forms a dip at $z_0=5$ in the well.  After this, amplitude of the soliton increases and recovers its original profile.   The width of the well is compressed when we increase the $\kappa$ value to $1.5$ which is demonstrated in Fig. \ref{fig5}(d).   
\subsubsection{Nonlinear tunneling with exponential background}
In optical communication systems, tunneling of solitons through the barrier/well plays a very significant role \cite{33}.  Motivated by this, we also investigate this special case by considering a nonlinear barrier or nonlinear well with an exponential background by choosing $\beta(z)$ and $R(z)$ in the form \cite{34}
\begin{equation}
\label{barrier2}
\beta(z)=\beta_0, \;\;\; R(z)= R_0 \exp(-r z)+h\; \sech^2{[\kappa (z-z_0)]},
\end{equation}
with $r$ is the decaying or increasing the parameter.

\begin{figure*}[!ht]
	\begin{center}
		\subfigure[]
		{
			\resizebox{0.4\textwidth}{!}{\includegraphics{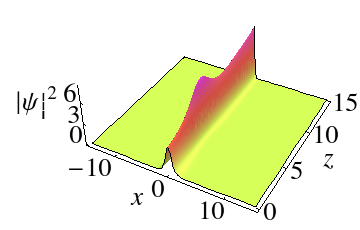}}
			\label{fig:1}
		}~~ 
		\subfigure[]
		{
			\resizebox{0.4\textwidth}{!}{\includegraphics{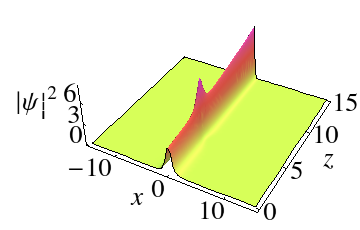}}
			\label{fig:2}
		}\\ \vspace{-0.3cm}
		\subfigure[]
		{
			\resizebox{0.4\textwidth}{!}{\includegraphics{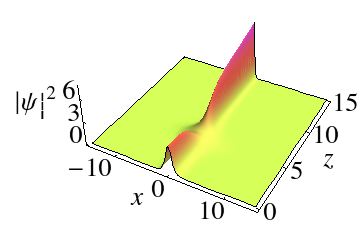}}
			\label{fig:3}
		}~~
		\subfigure[]
		{
			\resizebox{0.4\textwidth}{!}{\includegraphics{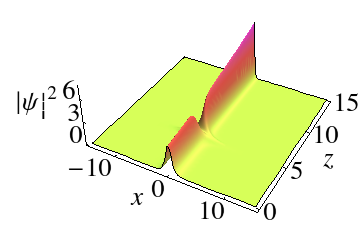}}
			\label{fig:4}
		}
	\end{center}
	\vspace{-0.3cm}
	\caption{Propagation of soliton through nonlinear barrier (a) and (b) for $h=1$ and nonlinear well (c) and (d) for $h=-1$.  The other parameters are $a=0.5,b=0.75, \beta_0=1.5$, $\chi_0=1$, $z_0=5$ with (a)  and (c) $\kappa=0.5$ and (b) and (d) $\kappa=1.5$. }	  	\label{fig6}
\end{figure*}

In Fig. \ref{fig6} we show the qualitative nature of nonlinear tunneling effect of the obtained optical soliton (\ref{cdis:eq6}) propagating through the nonlinear barrier/well (\ref{barrier2}).  We acquire the tunneling of optical soliton through nonlinear barrier as shown in Figs. \ref{fig6}(a)-(b) for $h=1$ and $r=-0.05$ in (\ref{barrier2}).   In Fig. \ref{fig6}(a), we see when the soliton propagates through the barrier, its amplitude enhances at $z_0=5$.  After reaching the barrier, the soliton recovers the original profile. Further, we notice that width of the soliton elongates along the direction of propagation at the barrier for a lower value of the barrier's width ($\kappa=0.5$) and for a higher value of $\kappa$ (=$1.5$), the width of soliton gets compressed at the barrier which is revealed in Fig. \ref{fig6}(b).  The propagation of soliton through the nonlinear well with the choice $h=-1$ is depicted in Figs. \ref{fig6}(c)-(d).  We notice that when the soliton propagates through the well, amplitude of pulse diminishes and constitute a dip at the well $z_0=5$ is shown in Fig. \ref{fig6}(c).  After this, amplitude of the soliton increases and recovers its original shape.  When we increase the $\kappa$ value to $1.5$ width of the well compresses which can be seen from Fig. \ref{fig6}(d). 

\subsection{Dispersion barrier/well}
In this subsection, we examine the tunneling characteristics of optical soliton profiles by considering with and without the exponential background of dispersion barrier/well.   
\subsubsection{Tunneling without exponential background}
To investigate the impact of tunneling behaviour of the obtained optical solitons, we consider the following dispersion barrier/well \cite{33,34}, that is
\begin{equation}
\label{barrier3}
\beta(z)=\beta_0+h \; \sech^2{[\kappa (z-z_0)]}, \;\;\; R(z)= R_0,
\end{equation}
where $h$, $\kappa$ and $z_0$ represents the height, width and location of the nonlinear barrier/well, respectively.  If $h$ is negative, $\beta(z)$ indicates the dispersion barrier and when $h$ is positive, $\beta(z)$ refers to the dispersion well.  
\begin{figure*}[!ht]
	\begin{center}
		\subfigure[]
		{
			\resizebox{0.4\textwidth}{!}{\includegraphics{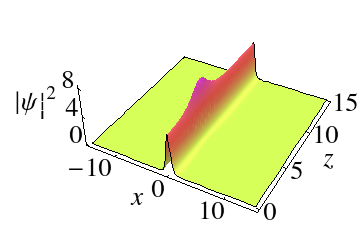}}
			\label{fig:1}
		}~~ 
		\subfigure[]
		{
			\resizebox{0.4\textwidth}{!}{\includegraphics{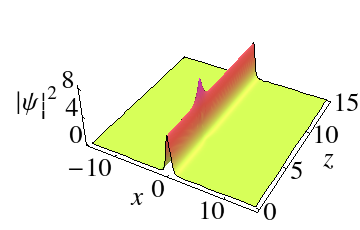}}
			\label{fig:2}
		}\\ \vspace{-0.3cm}
		\subfigure[]
		{
			\resizebox{0.4\textwidth}{!}{\includegraphics{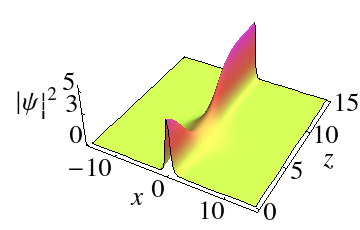}}
			\label{fig:3}
		}~~
		\subfigure[]
		{
			\resizebox{0.4\textwidth}{!}{\includegraphics{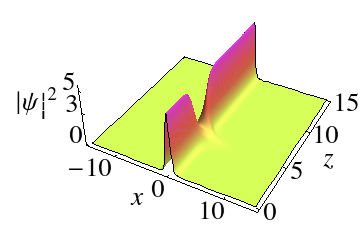}}
			\label{fig:4}
		}
	\end{center}
	\vspace{-0.3cm}
	\caption{Propagation of soliton through dispersion barrier (a) and (b) for $h=-0.35$ and dispersion well without an exponential background (c) and (d) for $h=4$.  The other parameters are $a=0.5,b=0.75, \beta_0=1.5$, $R_0=1$, $z_0=5$ with (a)  and (c) $\kappa=0.5$ and (b) and (d) $\kappa=1.5$. }
	\label{fig7}
\end{figure*}
\begin{figure*}[!ht]
	\begin{center}
		\subfigure[]
		{
			\resizebox{0.4\textwidth}{!}{\includegraphics{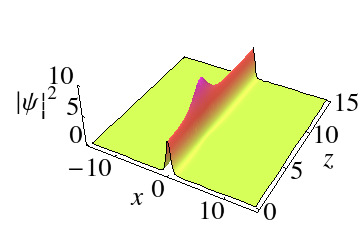}}
			\label{fig:1}
		}~~ 
		\subfigure[]
		{
			\resizebox{0.4\textwidth}{!}{\includegraphics{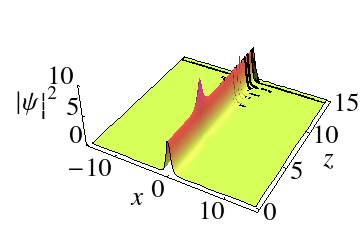}}
			\label{fig:2}
		}\\ \vspace{-0.3cm}
		\subfigure[]
		{
			\resizebox{0.4\textwidth}{!}{\includegraphics{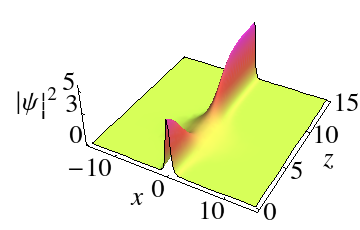}}
			\label{fig:3}
		}~~
		\subfigure[]
		{
			\resizebox{0.4\textwidth}{!}{\includegraphics{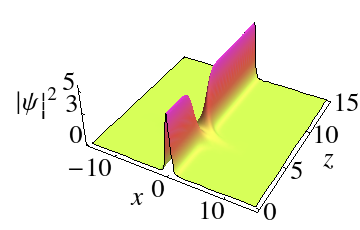}}
			\label{fig:4}
		}
	\end{center}
	\vspace{-0.3cm}
	\caption{Propagation of soliton through dispersion barrier (a) and (b) for $h=-0.35$ and $r=0.05$ and dispersion well with an exponential background (c) and (d) for $h=4$.  The other parameters are $a=0.5,b=0.75, \beta_0=1.5$, $R_0=1$, $z_0=5$ with (a) and (c) $\kappa=0.5$ and (b) and (d) $\kappa=1.5$. }	  	
	\label{fig8}
\end{figure*}
Figure \ref{fig7} represents the intensity profiles of the nonlinear tunneling effect of the obtained optical soliton (\ref{cdis:eq6}) propagating through the nonlinear barrier/well (\ref{barrier3}).  For $h=-0.35$ in (\ref{barrier3}), we observe that the optical solitons tunnels through nonlinear barrier as shown in Figs. \ref{fig7}(a)-(b)  Here we notice that when the soliton propagates through the barrier, its amplitude enhances at $z_0=5$ which is shown in Fig. \ref{fig7}(a) and then it regains to its original profile after it tunnels through the nonlinear barrier.  For a low value of the barrier's width ($\kappa=0.5$), at the barrier, we observe that the width of the soliton elongates along the propagation direction.  The width of soliton gets compressed at the barrier when we increase the value of $\kappa$ to $1.5$ which is demonstrated in Fig. \ref{fig7}(b) and attain its original profile after crossing over the barrier similar to the case as shown in Fig. \ref{fig7}(a).  For the choice $h=4$, the soliton propagates through nonlinear well as seen in Figs. \ref{fig7}(c)-(d).  In Fig. \ref{fig7}(c), we see amplitude of the pulse diminishes and forms a dip at the well $z_0=5$.  The amplitude of soliton increases and come back to its original profile after passing the well.   The width of well is compressed while we increase the $\kappa$ value to $1.5$ which is represented in Fig. \ref{fig7}(d).

\subsubsection{Tunneling with an exponential background}
To investigate this special case, we consider dispersion barrier or dispersion well on an exponential background in the form
\begin{align}
\label{barrier4}
\beta(z)&= \beta_0\exp(-r z)+h \; \sech^2{[\kappa (z-z_0)]}, \notag \\
 R(z)&= R_0\exp(-r z),
\end{align}
where $r$ is the decaying or enhancing the parameter.

In Fig. \ref{fig8}, we summarize the dynamical behaviour of the nonlinear tunneling effect of the obtained optical soliton (\ref{cdis:eq6}) propagating through the nonlinear barrier/well (\ref{barrier4}).  For $h=-0.35$ and $r=0.05$ in (\ref{barrier4}), we show the optical soliton tunnels through the dispersion barrier as shown in Figs. \ref{fig8}(a)-(b).  We observe that when the soliton propagates through the barrier, its amplitude enhances at $z_0=5$ as shown in Fig. \ref{fig8}(a).  After tunneling through the dispersion barrier, the optical soliton amplitude come back to its initial value along its direction of propagation.  The width of the soliton stretches along the direction of propagation at the barrier for the value of $\kappa=0.5$. We then increase the value of $\kappa$ to $1.5$, width of the soliton gets shortened at the barrier.  This result is illustrated in Fig. \ref{fig8}(b).  Next, we consider the choice $h=4$.  The optical soliton passes through dispersion well which can be visualized in Figs. \ref{fig8}(c)-(d).   We observe that the amplitude of optical pulse diminishes and forms a dip at the well $z_0=5$ as displayed in Fig. \ref{fig8}(c).  After passing the dispersion well, amplitude of the soliton increases and recovers its original profile.   In Fig. \ref{fig8}(d), we can visualize that the width of well is compressed while we vary the $\kappa$ value to $1.5$.  Our results confirms that after tunneling through the nonlinear and dispersion barrier/well the optical soliton keeps it amplitude and width unchanged all along its propagation distance. These theoretical findings may help experimentalist to realize the very stable optical soliton especially in exponentially varying background.
\section{Conclusion}
We have studied the inhomogenous soliton solutions of the NLS equation with distributed coefficients, namely dispersion, nonlinearity, tapering and $\mathcal{PT}$ - symmetric potential. We have investigated dynamical behaviours of the constructed soliton solutions under the influence of constant and exponentially distributed dispersion profiles.  From our investigations, we have found that the intensity of soliton enhances when we vary the strength of real and imaginary part of $\mathcal{PT}$ - symmetric Rosen-Morse potential.  On the other hand when we change the other parameters such as the strength of dispersion and nonlinear parameter we have observed that the width of the soliton stretches and compresses.  We have also investigated nonlinear tunneling effects of the obtained soliton profiles by considering two different types of nonlinear barrier/well and dispersion/well.  Our results show that the soliton's amplitude enhances and suppresses when we vary the height and width of the nonlinear barrier/well and dispersion barrier/well.  Our results will be helpful in their relative optical experiments.

\begin{acknowledgements}
KM wishes to thank the Council of Scientific and Industrial Research, Government of India, for providing the Research Associateship under the Grant No. 03/1397/17/EMR-II. The work of MS forms part of a research project sponsored by National Board for Higher Mathematics, Government of India, under the Grant No. 02011/20/2018NBHM(R.P)/R\&D 24II/15064.
\end{acknowledgements}

\end{document}